# A Non-Volatile Heterogeneous Quantum Dot III-V/Si DFB Laser with Optical Memristive Behavior

STANLEY CHEUNG,[1,†] BASSEM TOSSOUN,[2,†] DI LIANG,[3] YUAN YUAN[4] YINGTAO HU,[2] GEZA KURCZVEIL,[2] XUCHENG YANG,[1] AND RAYMOND BEAUSOLEIL[2]

[1] *North Carolina State University, Department of Electrical and Computer Engineering, 2410 Campus Shore Dr., Raleigh, NC. 27606, USA*
[2] *Hewlett Packard Labs, Large-Scale Integrated Photonics Laboratory, 820 N. McCarthy Blvd., Milpitas, CA. 95035, USA*
[3] *University of Michigan, Ann Arbor, Department of Electrical and Computer Engineering, 1301 Beal Ave., Ann Arbor, MI. 48109, USA*
[4] *Northeastern University, Department of Electrical Engineering, 500 MacArthur Blvd., Oakland, CA. 94613, USA*
[†]*These authors contributed equally.*
**scheung3@ncsu.edu*

**Abstract:** In this work, we introduce a non-volatile heterogeneous quantum dot (QD) III-V/$Al_2O_3$/Si distributed feedback (DFB) laser exhibiting optical memristive behavior. The device operates in the O-band (~1300 nm) with a threshold current density of 234 A/cm² and a side-mode suppression ratio exceeding 48 dB. Co-integrated $Al_2O_3$-based memristors produce bipolar resistive switching, yielding non-volatile wavelength shifts of ~ 46 pm and ~ 17 dB peak power contrast with zero static holding power. The III-V/$Al_2O_3$/Si heterojunction memristor I–V hysteresis is also modeled. This new device enables simultaneous coherent light generation and persistent optical state storage, establishing a new class of active photonic memory for neuromorphic and reconfigurable WDM applications.

## 1. Introduction

The rapid proliferation of data-intensive applications such as artificial intelligence, neuromorphic computing, and high-bandwidth optical communications has intensified the demand for large-scale photonic integrated circuits (PICs) that combine high-performance light sources with reconfigurable, non-volatile functionality [1–4]. Silicon photonics has emerged as the dominant platform for large-scale PIC integration, owing to its compatibility with complementary metal-oxide-semiconductor (CMOS) fabrication processes and its mature ecosystem for passive components [5]. However, silicon's indirect bandgap fundamentally limits its utility as a gain medium, necessitating the heterogeneous integration of III-V compound semiconductors to realize efficient, on-chip laser sources. Among the most compelling gain materials for heterogeneous III-V/Si lasers are self-assembled quantum dots (QDs), which offer a discrete, atom-like density of states that confers several distinct advantages over quantum well counterparts [6–9]. These include reduced threshold current density, enhanced temperature stability, lower linewidth enhancement factor, and superior tolerance to crystallographic defects — the latter being particularly significant in the context of direct epitaxial growth on silicon substrates [10]. QD-based lasers have demonstrated remarkable performance as heterogeneously bonded sources on silicon, and recent advances in wafer bonding and die-to-wafer integration have enabled the co-fabrication of QD III-V gain sections with silicon waveguide architectures at wafer scale [5,11–17]. Distributed feedback (DFB) lasers, which leverage periodic Bragg grating structures to enforce single-longitudinal-mode oscillation, are indispensable for coherent optical communications, wavelength-division multiplexing (WDM), and sensing applications requiring narrow linewidth and stable emission wavelength [13,18]. The realization of heterogeneous QD III-V/Si DFB lasers thus represents a convergence of the wavelength selectivity demanded by these applications with the low-noise, thermally robust characteristics intrinsic to QD gain media [13,19,20].

A particularly transformative and largely unexplored frontier lies at the intersection of integrated photonics and neuromorphic hardware: the incorporation of non-volatile, reconfigurable optical functionality directly embedded within the laser cavity [21,22]. Memristive behavior - characterized by a hysteretic, history-dependent relationship between applied stimulus and device state - has been extensively studied in electronic resistive switching devices for in-memory computing and synaptic emulation [23–30]. The translation of this paradigm to the optical domain, wherein a photonic device exhibits persistent, non-volatile switching between discrete emission states without continuous power consumption to maintain those states, opens transformative possibilities for optical memory elements, photonic synapses, and reconfigurable laser networks [21,31–36,36–40].

In this work, we present a non-volatile heterogeneous quantum dot III-V/$Al_2O_3$/Si DFB laser (Fig. 1a-c) exhibiting optical memristive behavior - a device architecture that, to the best of our knowledge, has not been previously demonstrated. By engineering the interplay between the QD gain medium, the silicon waveguide platform, and the DFB grating structure, we demonstrate that the laser can be reversibly switched between stable, non-volatile emission states through controlled electrical stimuli, with the device retaining its programmed state in the absence of a holding bias. We characterize the non-volatile wavelength switching, power contrast ratios, and switching dynamics inherent in the observed memristive response. We also accurately model, for the first time, the electrical hysteresis of the III-V/$Al_2O_3$/Si heterojunction memristors. These results establish a new class of active photonic memory devices and lay the groundwork for their integration into large-scale neuromorphic photonic circuits and reconfigurable WDM systems.

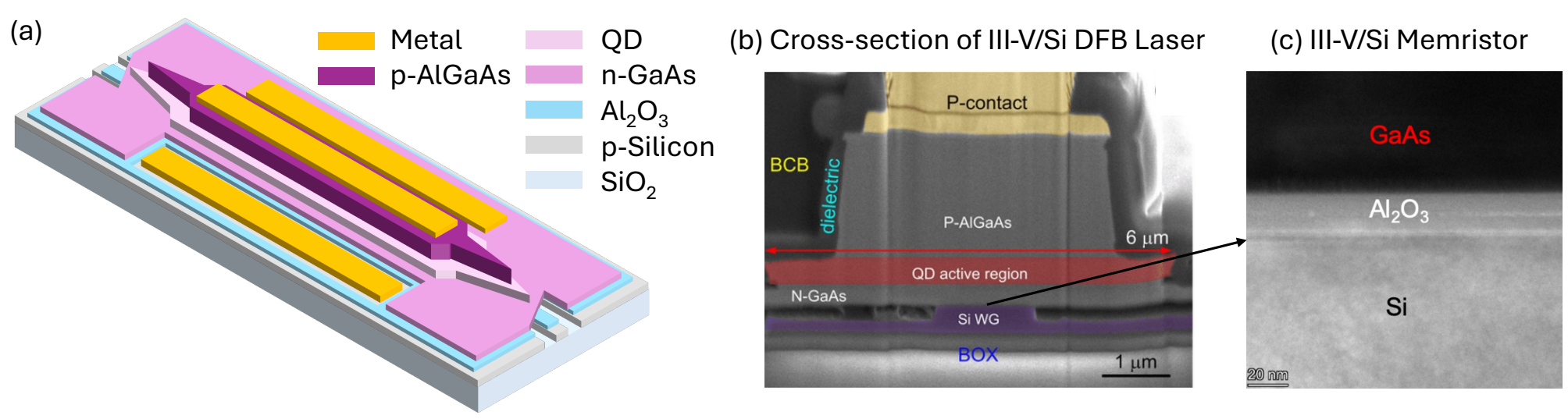

Fig. 1 (a) 3D schematic of III-V/$Al_2O_3$/Si memristive DFB laser. Cross-sectional images of (b) hybrid laser structure and (c) co-integrated III-V/$Al_2O_3$/Si memristor at the interface of the hybrid waveguide.

## 2. Fabrication

The fabrication process for realizing non-volatile memristive memory on a heterogeneous III-V/Si QD platform is outlined in Fig. 2. In-house device fabrication starts with a 100 mm SOI wafer that consists of a 300 nm top silicon layer and a 2 μm thick buried oxide (BOX) layer. A blanket ion implantation of boron was performed (4E+16 $cm^{-3}$) to facilitate conductivity of integrated volatile and non-volatile semiconductor-insulator-semiconductor capacitive (SISCAP) phase shifters . Alignment marks and grating couplers were both patterned using a 248 nm KrF ASML DUV stepper and etched 145 nm with $Cl_2$-based gas chemistry. A 9-step boron implantation scheme was used to create p++ contacts (1E+20 $cm^{-3}$) for the SISCAP phase shifters . Next, silicon rib waveguides and vertical outgassing channels (VOCs) were patterned and etched 217 nm and 300 nm respectively with laser end-point detection. The distributed feedback gratings are then patterned with a period of 394 nm at a 50 % duty cycle with an etch depth of 10 nm. The silicon wafer is then thoroughly cleaned with a Piranha solution followed by a short dilute hydrofluoric (HF) acid etch to remove any residual hard masks. An $O_2$ plasma clean was then performed followed by a SC1 and SC2 clean. The III-V QD epitaxial wafer is cleansed with acetone, methanol, and IPA, followed by an $O_2$ plasma clean and a $NH_4OH$:$H_2O$ (1:10) dip for 1 min. Next, a 10 nm thick dielectric layer of $Al_2O_3$ is deposited on both the III-V and SOI wafer via atomic layer deposition (ALD) at 300 °C. The two samples are then mated manually at room temperature using a Finetech flip-chip bonder and then wafer-bonded under pressure for 300 °C (2 hour ramp) for a total of 15 hours. After wafer-bonding, the backside of the III-V was mechanically lapped until a 100 µm thickness of III-V was left. Next, the p-GaAs substrate is removed using a wet etch which selectively stops on a 20 nm p-AlGaAs layer. The etch stop layer is then subsequently removed using buffered HF acid to reveal a clean 100 nm thick p-GaAs contact layer. Metal contacts consisting of Pt/Ti/Pt/Au (5/25/50/250 nm) were deposited onto the p-GaAs as the laser p-contact metal. III-V mesas are then defined by etching using a SiN hard mask and an ICP etcher using a $Cl_2$-based gas chemistry stopping within a 100 nm n-AlGaAs etch stop layer using laser endpoint detection. This etch stop layer is then wet etched to reveal a clean n-GaAs contact. A combination of Pd/Ge/Ti/Au/Ti (30/60/50/200/10 nm) metals were deposited to create the laser n-contact metal and annealed at 300 °C for 30 seconds. Next, the III-V QD mesas were isolated by selectively dry etching various regions of the n-GaAs and ALD dielectric. Next, a plasma enhanced chemical vapor deposition (PECVD) SiN cladding was deposited followed by a thick BCB layer to minimize electrical parasitics. Finally, a thick layer of Ti/Au (1.6 µm) was evaporated to create metal probe pads for both laser p and n-contacts.

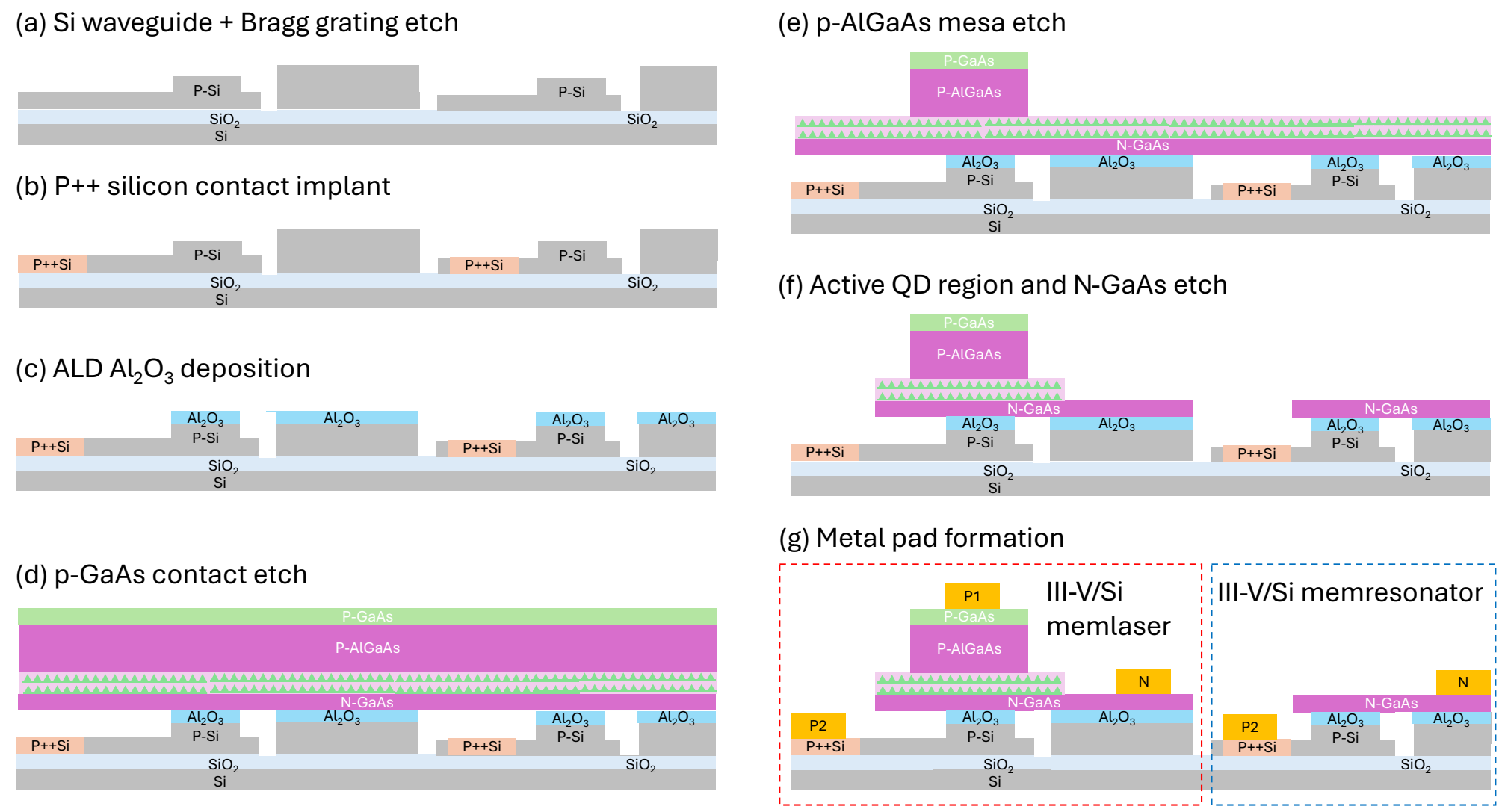

Fig. 2. Simplified schematic of fabrication process for a heterogeneous III-V/dielectric/Si QD platform with embedded non-volatile memristive memory.

## 3. Device Characterization

### 3.1 III-V/$Al_2O_3$/Si Non-Volatile Memristor Heterojunction

To confirm electrical memristive behaviour of the III-V/$Al_2O_3$/Si structure, we performed I-V measurements on hybrid ring resonator structures with cross-sections indicated in Fig. 3a. This test structure consists of a ring radius = 25 µm with a waveguide width = 0.8 µm for a total area of 123.65 $\mu m^2$. Cross-sectional electron microscopy (Fig. 3a) confirms a well-defined layer stack comprising a n-GaAs epi-layer interfaced with a 23 nm $Al_2O_3$ tunnelling oxide atop a patterned p-Si waveguide. The air gap ensures electrical isolation between the highly conductive n-GaAs and p-Si regions. The semi-logarithmic current–voltage (*I*–*V*) sweeps acquired over five consecutive cycles (Fig. 3b) demonstrate reproducible bipolar resistive switching behavior, wherein the "Set" process is activated at a negative bias (~ − 5 to − 7 V) and the "Reset" process occurs at positive bias (~ + 3 to + 5 V), yielding an on/off current ratio exceeding one order of magnitude (~10×) at low read voltages. An entire cycle is defined as 0 V → − 7.5 V → 0 V → 7.5 V → 0 V with a 1 mA current compliance applied during measurement. The corresponding resistance–voltage (*R*–*V*) characteristics (Fig. 3c) corroborate this switching polarity, revealing a high-resistance state (HRS) approaching ~35 kΩ and a low-resistance state (LRS) of approximately 2 – 5 kΩ, with the resistance hysteresis window remaining stable across cycling. The cycle-to-cycle variability observed in both the *I*–*V* and *R*–*V* curves — represented by the spread among the colored traces — indicates moderate stochastic fluctuations inherent to filamentary or interface-trap-mediated conduction mechanisms within the ultrathin $Al_2O_3$ dielectric, a phenomenon commonly attributed to the probabilistic nature of conductive filament formation within the bulk $Al_2O_3$ and rupturing at the n-GaAs/oxide and p-Si/oxide interfaces. Collectively, these results establish the viability of monolithically integrated III-V/Si memristors as a promising platform for non-volatile memory and neuromorphic computing applications. It should be noted that after 5 cycles, the I-V hysteresis started exhibiting permanent conductivity behaviour without the possibility to reset. Future work is on-going to explore pulsed biasing conditions to minimize these failure rates.

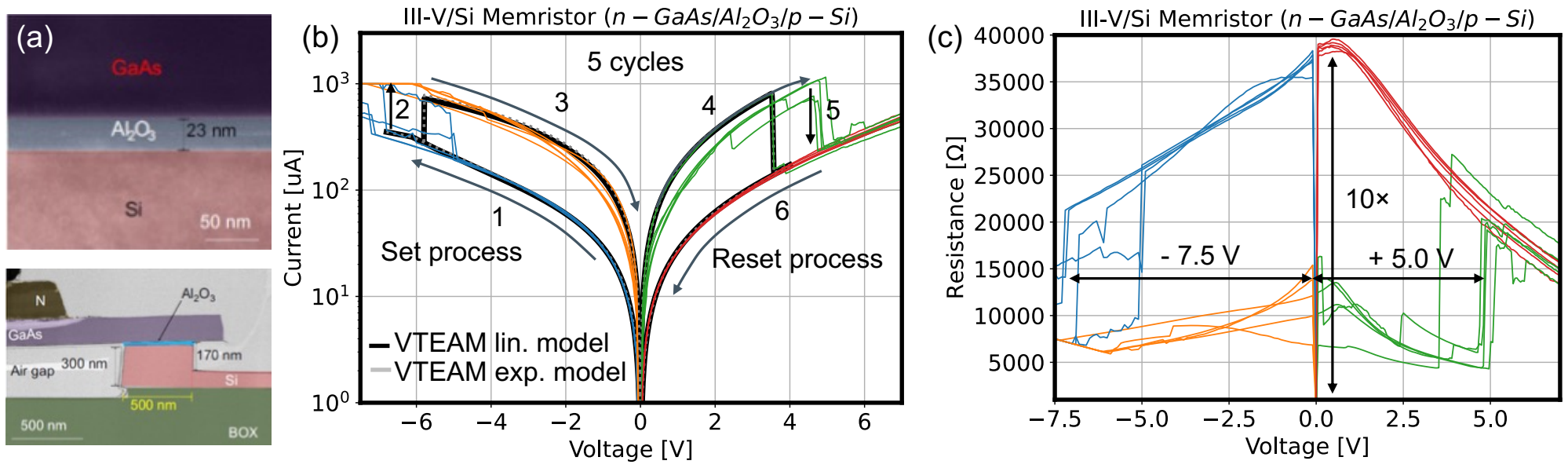

Fig. 3 (a) Cross sectional image of III-V/$Al_2O_3$/Si optical memristor and integration within a waveguide, (b) measured I-V hysteresis for 5 cycles indicating electrical non-volatility, (c) non-volatile resistance.

To the best of our knowledge, there has been no reported work in modeling III-V/insulator/Si heterojunction memristors. The current–voltage (I–V) hysteresis characteristics of the n-GaAs/$Al_2O_3$/p-Si heterojunction memristor can be modelled using the Voltage Threshold Adaptive Memristor (VTEAM) framework [41] by the following:

$$\frac{dw(t)}{dt} = \begin{cases} k_{off} \cdot \left(\frac{v(t)}{v_{off}} - 1\right)^{\alpha_{off}} \cdot f_{off}(w), & 0 < v_{off} < v \\ 0, & v_{on} < v_{off} < v \\ k_{on} \cdot \left(\frac{v(t)}{v_{on}} - 1\right)^{\alpha_{on}} \cdot f_{on}(w), & v < v_{on} < 0 \end{cases} = f(w, v) \tag{1}$$

$$\frac{i(t)}{v(t)} = G(w, v) = \begin{cases} \left[R_{ON} + \frac{R_{OFF} - R_{ON}}{w_{off} - w_{on}}(w - w_{on})\right]^{-1} & (linear\ fit) \\ \frac{e^{-\frac{(R_{OFF}/R_{ON})}{w_{off} - w_{on}}(w - w_{on})}}{R_{ON}} & (exponential\ fit) \end{cases} \tag{2}$$

where $w(t)$ is the memristor internal state variable, $v(t)$ is the voltage across the memristive device, $i(t)$ is the current passing through the memristor, $G(w, v)$ is the device conductance, and $t$ is time. The VTEAM model governs the time evolution of an internal state variable $w \in [w_{on}, w_{off}]$ through a piecewise ordinary differential equation with positive and negative voltage thresholds $v_{off}$ and $v_{on}$, power-law exponents $\alpha_{off}$ and $\alpha_{on}$, and rate constants $k_{off}$ and $k_{on}$, with an ideal rectangular window function enforcing hard boundary conditions. The determination of G(w, v) depends on the fitting of either a linear or exponential model as shown in Eq. 2. Two distinct current–voltage relationships were employed depending on the physical conduction regime: an ohmic model, $i = |v| \cdot G(w)$, was used for the negative-bias sweeps, where point-by-point resistance analysis confirmed that conduction is dominated by a linearly state-modulated resistance varying between approximately 35.5 kΩ (OFF state) and 2.7 kΩ (ON state); and a diode-like model, $i = G(w) \cdot R^{ON} \cdot I_0 \cdot (\exp(|v|/nV_T) - 1)$, was applied to the forward-bias sweeps to capture the rectifying heterojunction character. Both linear and exponential conductance interpolation schemes for $G(w)$ were evaluated as shown in Fig. 3b indicated by the black and gray colored curves. Parameter optimization was performed using a two-stage procedure: a global search via differential evolution followed by local refinement with the Nelder–Mead simplex method, minimizing a log-scale root-mean-square error objective to appropriately weight the multi-decade current range. The best-fit VTEAM parameters can be found in the Supplementary section. The negative-bias return sweep (−8 → 0 V, ON state) was reproduced with high fidelity (log-RMS = 0.039), consistent with a well-defined low-resistance conduction path. The forward SET sweep exhibited a gradual resistance reduction from 35.5 to 21.5 kΩ between 0 and −6.0 V — attributed to progressive VTEAM state evolution active from a threshold of approximately −3.5 V — followed by an abrupt switching event at −6.1 V, where the resistance dropped discontinuously to 2.7 kΩ.

### *3.2 III-V/$Al_2O_3$/Si Non-Volatile Memristive DFB Laser*

The III-V/$Al_2O_3$/Si QD DFB memristive laser, as shown in Fig. 4a consists of a silicon waveguide with height = 300 nm, etch depth = 170 nm, and varying silicon waveguide widths along the propagation direction. The III-V/Si taper transition, as shown in the inset of Fig. 4a, starts with a 50 μm long taper and widens the silicon waveguide width from $w_{Si}$ = 0.5 μm to 2.0 μm. Simultaneously, the III-V waveguide is introduced via an inverse taper of length $L_{taper}^{III-V}$ = 25 μm, over which the III-V width expands from $w_{III-V}$ = 1.30 μm to $w_{III-V}$ = 4.0 μm, while the silicon width beneath it is maintained at $w_{Si}$ = 2.0 μm. This counter-tapered geometry enables a gradual, low-loss transfer of optical power from the high-index-contrast Si waveguide mode into the broader III-V active region mode, where the QD confinement factor $\Gamma_{QD}$ reaches approximately 13% at a III-V active width of 5 μm and Si width of 1.5 μm. After the III-V/Si transition region, the silicon waveguide width is tapered down from $w_{Si}$ = 2.0 μm to 1.5 μm with a taper length of $L_{taper}^{Si}$ = 50 μm. The optical transition from active to passive waveguides was theoretically determined to be < 0.4 dB loss with ~ -38 dB residual reflections from 3D-FDTD simulations.

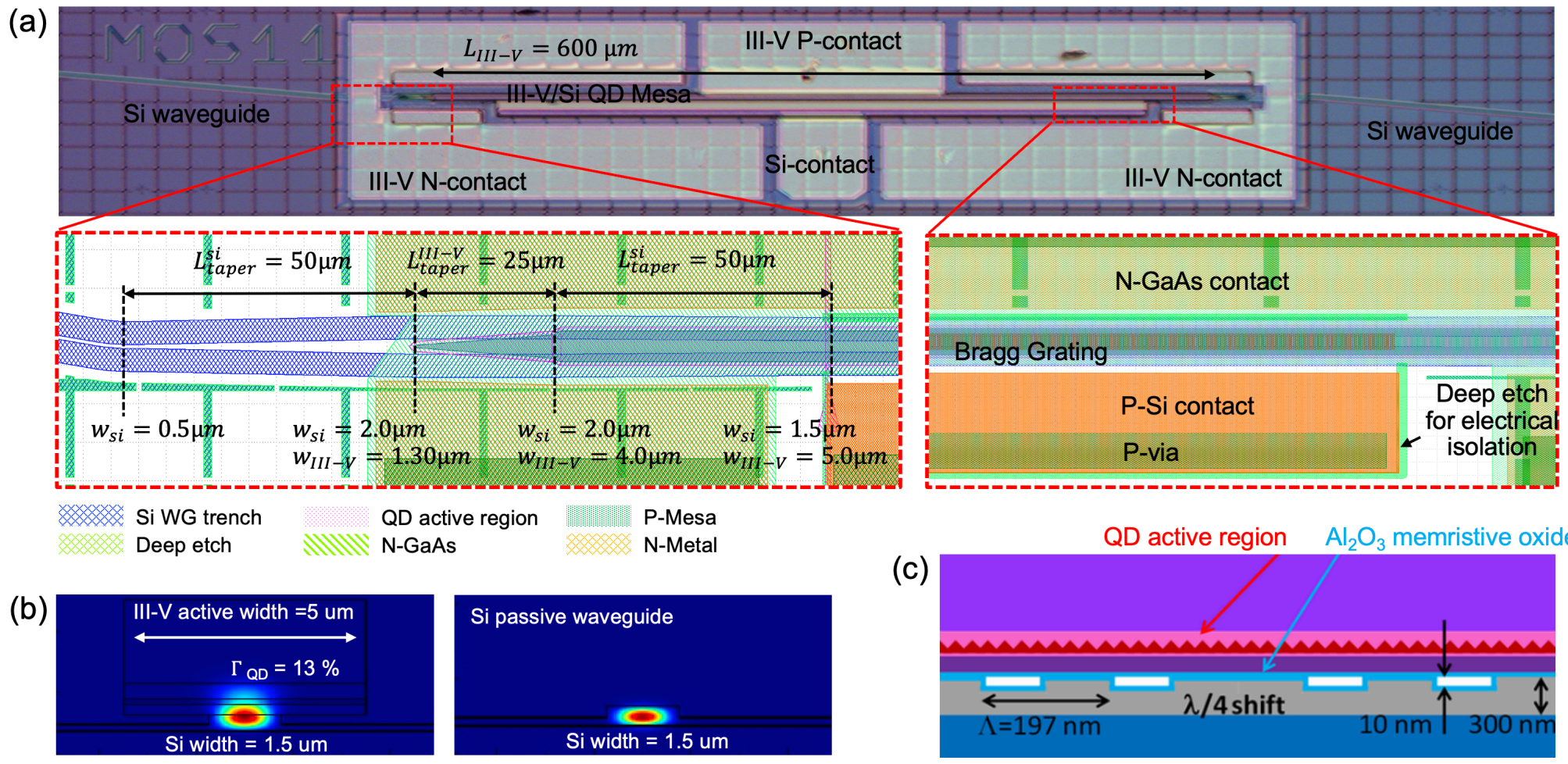

Fig. 4. (a) Microscope image of quantum dot (QD) III-V/$Al_2O_3$/Si DFB laser with key design parameters such as the III-V/Si taper transition and contact regions for the co-integrated memristor. (b) Optical mode simulations in the active and passive regions. (c) Side-view schematic of central DFB region with design parameters.

The DFB gratings are defined by e-beam lithography and consist of a pitch = 197 nm with 50% duty cycle and 10 nm etch depths as shown in Fig. 4c. The cavity center includes a λ/4 shift such that single-mode lasing occurs at 1310 nm for a total DFB cavity length of 499.594 µm. After silicon processing, a thin layer of 11 nm ALD $Al_2O_3$ was deposited on both silicon and III-V regions such that covalent bonding of both pieces will result in a 22 nm thick GaAs/$Al_2O_3$/Si capacitor. The III-V p-mesa width is 4 µm wide and the QD active region below it is 5 µm which consists of a proprietary design of 5 InAs/GaAs QDs. The QD active region is intentionally wider to prevent any undercutting from wet etch processing. Mode simulations in Fig. 4b suggests a quantum dot active confinement factor of $\Gamma_{QD}$ = 13 %. Below the active region is a large n-layer to facilitate ground contacts. Metal contact pads are formed on the p-GaAs, n-GaAs, and p-Si regions where the n-GaAs/p-Si contacts facilitate either high-speed volatile phase tuning or non-volatile memristive effects.

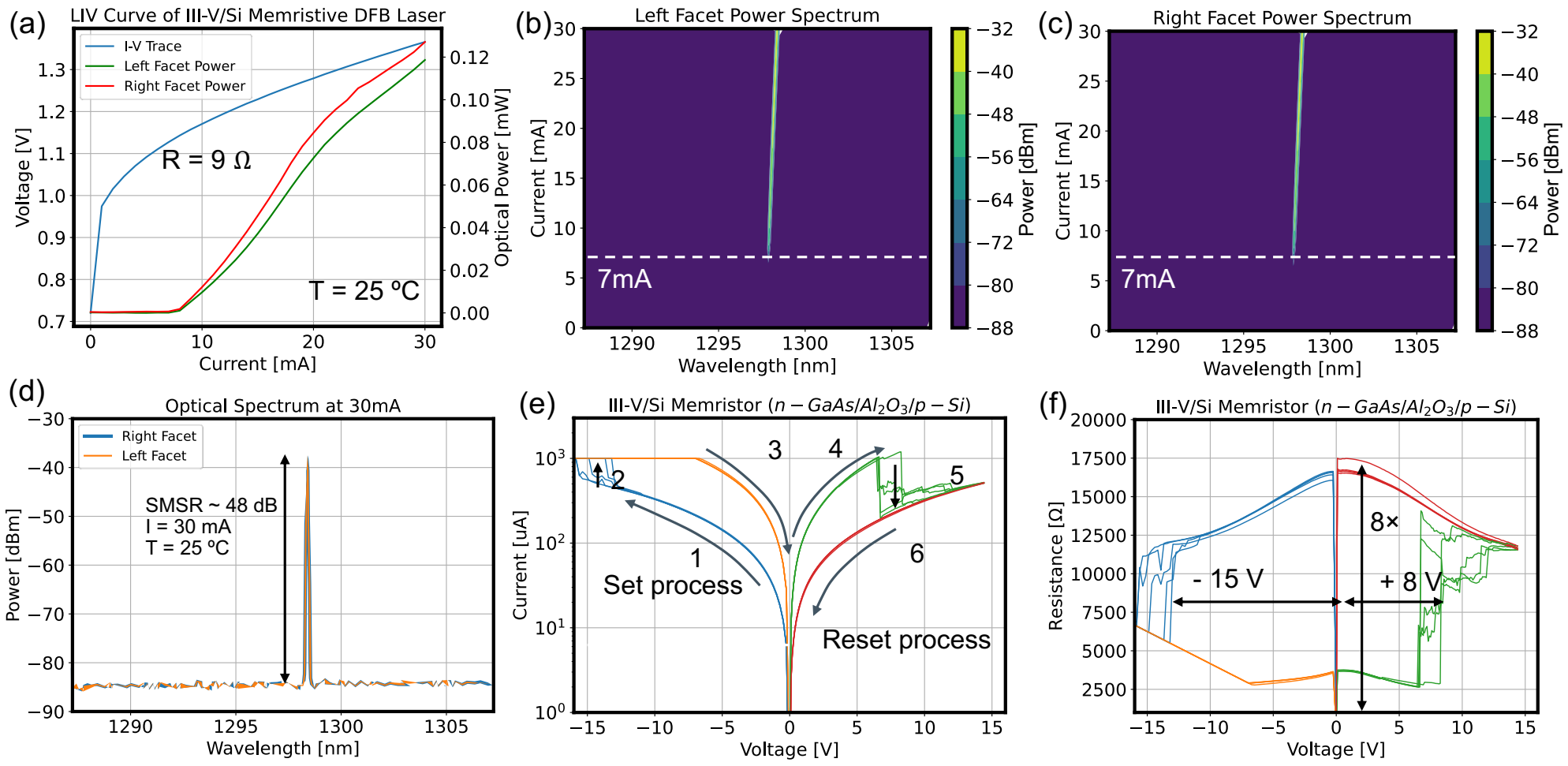

Fig. 5 (a) Continuous wave (CW) light-current-voltage (LIV) curves. Measured optical spectra for (b) left output facet, and (c) right output facet. (d) Observed side-mode-suppression-ratio (SMSR) ~ 48 dB. III-V/Si non-volatile memristive testing indicating (b) I-V hysteresis for 5 cycles and corresponding (c) non-volatile resistance.

Fig. 5 presents an initial optoelectronic characterization of the monolithically integrated III-V/$Al_2O_3$/Si memristive DFB laser, demonstrating the simultaneous co-existence of coherent light emission and non-volatile resistive switching within a single heterojunction device. The light-current-voltage (*LIV*) characteristics (Fig. 5a), measured at T = 25 °C on a temperature controlled stage, reveal a well-defined lasing threshold with symmetric optical output from both facets, a series resistance of R = 9 Ω and a maximum optical power output approaching ~ 0.12 mW at 30 mA injection current. Output optical powers are normalized to a 13 dB coupling loss determined from spiral waveguide test structures. Optical power was collected from both ends of the laser with a pair of cleaved SMF-28 optical fibers vertically angled at 7°. This coupling angle is far from optimal due to mechanical constraints and we expect to see much improved coupling losses in the future with the correct fiber angles. Spectrally resolved power maps acquired from the left and right facets (Fig. 5b–c) confirm the onset of single-mode lasing above a threshold current of approximately 7 mA, with the emission wavelength anchored at ~ 1300 nm by the DFB grating — a telecom-relevant O-band wavelength — and exhibiting strong longitudinal and transverse mode suppression that persists across the full current range and spectrum of 20 nm. The device exhibits a wavelength to current shift of 22.5pm/mA and threshold current density of 234 A/cm$^2$. At an injection current of 30 mA, the optical spectrum (Fig. 5d) yields a side-mode suppression ratio (SMSR) of ~ 48 dB, confirming robust single-frequency operation suitable for coherent communication applications. Concurrently, the same device exhibits bipolar resistive switching in as evidenced in the *I–V* characteristics (Fig. 5e), with set and reset transitions occurring at negative and positive bias polarities, respectively, and an on/off current ratio spanning nearly three orders of magnitude. The corresponding *R*–*V* hysteresis curves (Fig. 5f) resolve a high-resistance state (HRS) of ~17.5 kΩ and a low-resistance state (LRS) below ~2.5 kΩ, sustaining a resistance modulation window exceeding 7× across repeated cycles. The seamless integration of DFB lasing at 1300 nm with memristive switching functionality within a single *n*-GaAs/$Al_2O_3$/*p*-Si platform represents a significant advance toward optoelectronic-in-memory architectures capable of simultaneously processing and transmitting information at the physical layer.

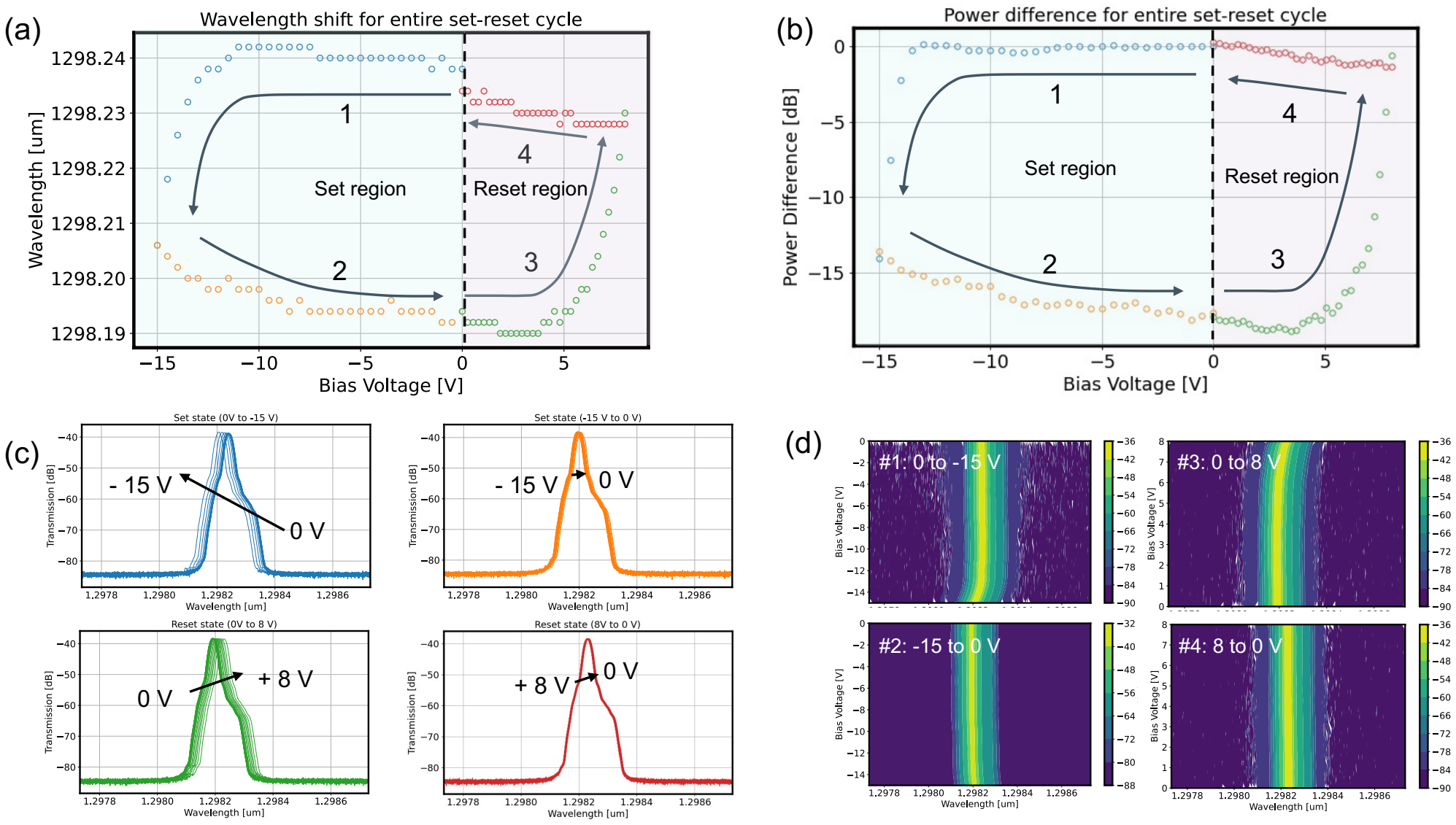

Fig. 6. Evolution of (a) peak wavelength shift and (b) power difference over the entire electrical set-reset cycle. (c) Measured optical spectra for complete set-reset cycle. (d) Spectral evolution vs. bias voltage indicating non-volatile wavelength shift of ~ 46 pm.

In order to investigate non-volatile optical memory functionality of the QD DFB laser, we perform the same I-V cycle test as shown in Fig. 5e and track the peak lasing wavelength while maintaining a laser drive current of 10 mA. Fig. 6a shows the evolution of the peak wavelength shifts over the entire electrical set-reset cycle. Applying a bias from 0 to -15 V (blue data points) results in a 46 pm non-volatile wavelength shift. Non-volatility is verified by ramping the bias down from -15 to 0 V (orange data points) where the wavelength shift remains permanently blue-shifted without any bias. Non-volatility is believed to be due to filamentation induced charge trap regions which serve to alter the electric field across the oxide and cause a refractive index change due to carrier re-distribution. By applying a positive bias up to 7 V (green data points), the filamentation is ruptured and is associated with a red shift. The rupturing of the filamentation is known to be associated with joule heating and consistent with observed wavelength red shifting. Ramping back down to 0 V nearly resets the peak wavelength to its original position (red data points). The reset is not completely perfect due to possible degradation of the $Al_2O_3$ film. The spectral evolution of the entire set-reset cycle is shown in Fig. 6c-d. The original lasing wavelength peak occurs at 1298.24 nm with a current bias of 10 mA and the non-volatile wavelength shift of 46 pm results in a 16.52 dB amplitude difference. If non-volatile memristive lasers were to be used as non-volatile elements, retention times are of interest. Although, retention times are not shown in this paper, we have in the past investigated a similar memristive mechanism in MZIs and de-interleaver filters, albeit with an alternating pair of $HfO_2$/$Al_2O_3$ dielectric stack [31]. Non-volatile state retention lasted > 24 hours and we expect similar values for this work. The non-volatile wavelength shift can be defined as $\Delta\lambda_{non\text{-}volatile} = (\lambda_{res}/n_{eff})\Delta n_{eff}$. Taking $n_{eff} = 3.337$ from mode simulations in Fig. 4b and $\lambda_{res}$ = 1298.24 nm at a drive current of 10 mA from Fig. 6b, the change in effective index $\Delta n_{eff}$ is estimated to be 1.6E-4.

The non-volatile switching speed and energy was evaluated on an alternate sample similar to the III-V/Si DFB structure described above. As shown in Fig. 7b A Keysight B1500A semiconductor parameter analyzer was used to generate high speed voltage pulses (50% duty cycle with 50 μs pulses) which are applied to the contacts between the p-Si and n-GaAs. A 100 kΩ resistor is placed in series with the device under test such that the device current can be measured using the sampling oscilloscope. The output optical signal from the DFB laser is fed into a tunable optical filter and then split into an OSA and PDFA + photodetector for time resolved non-volatile switching. In order to observe a non-volatile wavelength shift, twenty 50 μs pulses were needed for a total switching time of ~ 2 milliseconds. The photodetector signal (blue curve) in Fig. 7a indicates a non-volatile optical signal after the source voltage (green curve) is turned off, thus demonstrating wavelength shifts with zero static power consumption.

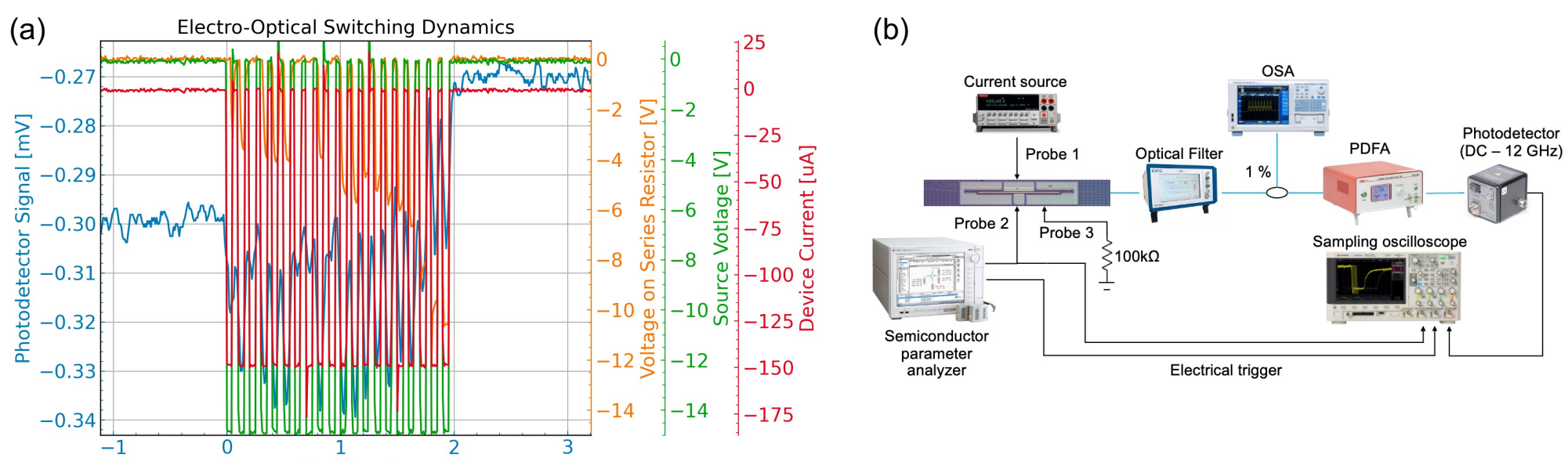

Fig. 7. (a) Electro-optical time dynamics for non-volatile wavelength shifting. (b) experimental setup.

These early devices did not survive long enough to perform multi-bit experiments, but it should be noted that thousands of conductance levels have been demonstrated using electrical $Al_2O_3$:$TiO_x$/$HfO_2$ memristors [26]. By leveraging these rapid developments, high bit resolution memristive lasers are a possibility with ultimate limiting factors due to finite laser linewidths or particular cavity designs.

## 4. Conclusion

In this work, we have demonstrated the first non-volatile heterogeneous quantum dot III-V/Si DFB laser exhibiting optical memristive behavior — a device that simultaneously integrates coherent, single-mode light generation with persistent, history-dependent resistive switching within a single monolithic heterojunction platform. The fabricated III-V/$Al_2O_3$/Si QD DFB laser operates at the telecom O-band (~1300 nm) with a threshold current density of 234 A/cm² and a SMSR exceeding 48 dB, confirming robust single-frequency operation. The co-integrated memristor, formed at the n-GaAs/$Al_2O_3$/p-Si interface, exhibits bipolar resistive switching with an HRS/LRS resistance ratio exceeding 7× and an on/off current ratio of approximately one order of magnitude. By performing I–V cycling while maintaining laser drive current, we demonstrated non-volatile wavelength shifts of ~46 pm accompanied by a ~17 dB peak power contrast — states that are retained without any static holding bias, corresponding to an estimated change in effective refractive index of $\Delta n_{eff} \approx 1.6\times10^{-4}$. The non-volatile switching mechanism is attributed to filamentary conductive path formation and rupture within the ultrathin $Al_2O_3$ dielectric, which redistributes carriers and modifies the refractive index experienced by the guided optical mode. Time-resolved electro-optical measurements further confirm that non-volatile wavelength shifts are achievable using ~2 ms pulsed electrical stimuli with zero static power consumption to maintain the programmed state. The I–V hysteresis characteristics of the III-V/$Al_2O_3$/Si memristive heterojunction were modeled using the VTEAM framework, with both ohmic and diode-like conduction regimes captured through linear and exponential conductance interpolation schemes, respectively. The model parameters were optimized via a two-stage global-local fitting procedure and reproduce the measured switching behavior with high fidelity across the multi-decade current range.
Taken together, these results address a fundamental challenge facing modern computing architectures — the von Neumann bottleneck — by demonstrating that memory and optical signal generation can co-exist within a single, CMOS-compatible photonic device. Unlike conventional approaches that treat the laser and memory element as separate components linked by interconnects, our architecture embeds non-volatile state storage directly within the laser cavity, enabling simultaneous opto-electronic in-memory processing at the physical layer with negligible idle power overhead. Future work will focus on improving $Al_2O_3$ dielectric reliability to extend endurance beyond five switching cycles, optimizing pulsed biasing schemes to suppress filament-induced device degradation, and scaling the architecture toward arrayed WDM configurations in which each channel of a DFB laser array is independently addressable as a non-volatile photonic memory element. The demonstrated device establishes a new class of active photonic memory and lays a critical foundation for their integration into large-scale neuromorphic photonic circuits and reconfigurable optical interconnect systems [39,42–44].

## 5. Back matter

### *5.1 Funding*

**Funding.** DOE ARPA-E ULTRALIT contract No. DEAR0001039.

### *5.2 Acknowledgment*

**Acknowledgment.** We thank funding from DOE ARPA-E ULTRALIT contract No. DEAR0001039.

### *5.3 Disclosures*

The authors declare no conflicts of interest.

### *5.4 Data availability statement*

**Data availability.** Data underlying the results presented in this paper are not publicly available at this time but may be obtained from the authors upon reasonable request.

### *5.5 Supplementary Document*

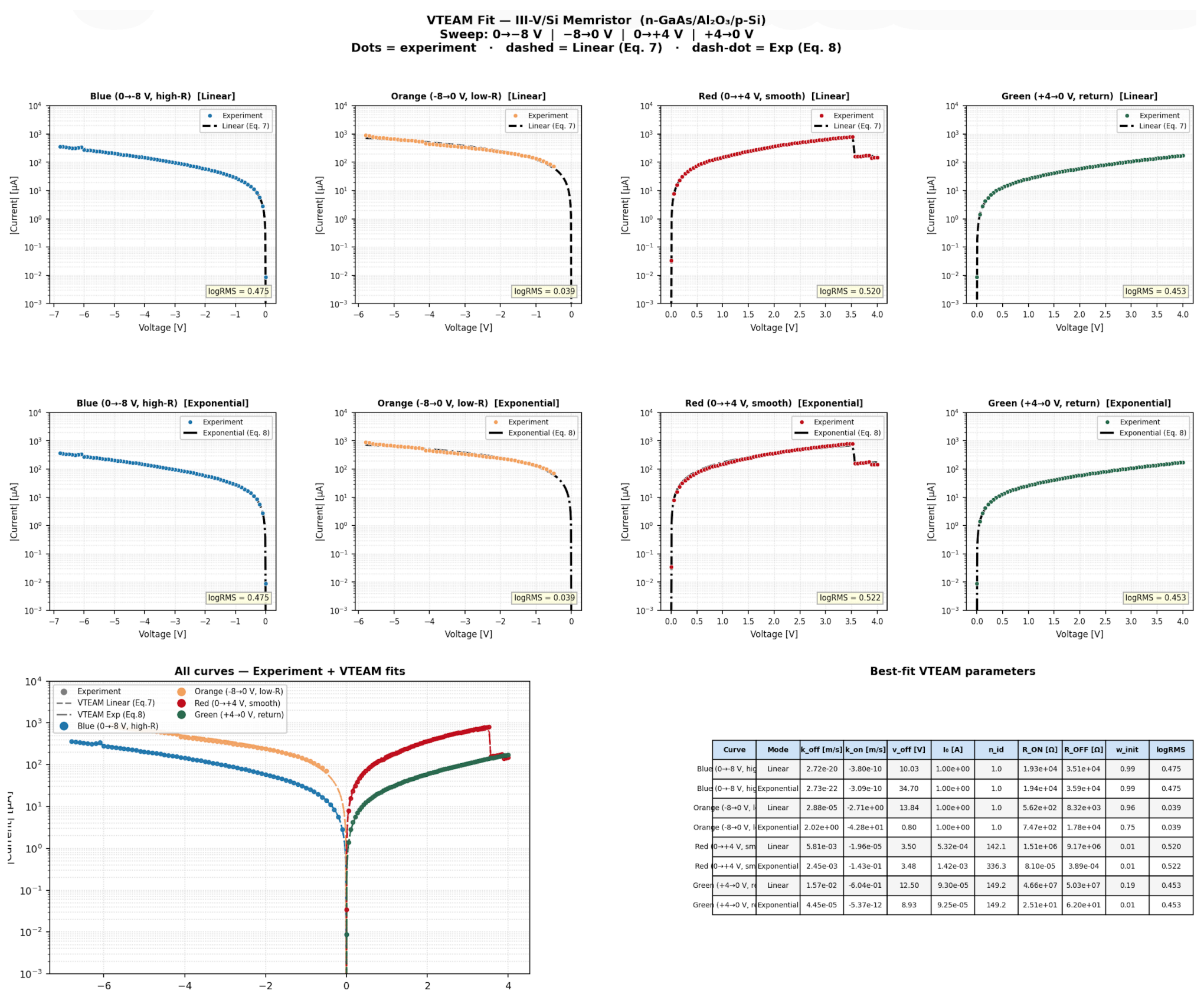

| Curve | Mode | k_off [m/s] | k_on [m/s] | v_off [V] | I₀ [A] | n_id | R_ON [Ω] | R_OFF [Ω] | w_init | logRMS |
|---|---|---|---|---|---|---|---|---|---|---|
| Blue (0→-8 V, hig | Linear | 2.72e-20 | -3.80e-10 | 10.03 | 1.00e+00 | 1.0 | 1.93e+04 | 3.51e+04 | 0.99 | 0.475 |
| Blue (0→-8 V, hig | Exponential | 2.73e-22 | -3.09e-10 | 34.70 | 1.00e+00 | 1.0 | 1.94e+04 | 3.59e+04 | 0.99 | 0.475 |
| Orange (-8→0 V, l | Linear | 2.88e-05 | -2.71e+00 | 13.84 | 1.00e+00 | 1.0 | 5.62e+02 | 8.32e+03 | 0.96 | 0.039 |
| Orange (-8→0 V, l | Exponential | 2.02e+00 | -4.28e+01 | 0.80 | 1.00e+00 | 1.0 | 7.47e+02 | 1.78e+04 | 0.75 | 0.039 |
| Red (0→+4 V, sm | Linear | 5.81e-03 | -1.96e-05 | 3.50 | 5.32e-04 | 142.1 | 1.51e+06 | 9.17e+06 | 0.01 | 0.520 |
| Red (0→+4 V, sm | Exponential | 2.45e-03 | -1.43e-01 | 3.48 | 1.42e-03 | 336.3 | 8.10e-05 | 3.89e-04 | 0.01 | 0.522 |
| Green (+4→0 V, re | Linear | 1.57e-02 | -6.04e-01 | 12.50 | 9.30e-05 | 149.2 | 4.66e+07 | 5.03e+07 | 0.19 | 0.453 |
| Green (+4→0 V, re | Exponential | 4.45e-05 | -5.37e-12 | 8.93 | 9.25e-05 | 149.2 | 2.51e+01 | 6.20e+01 | 0.01 | 0.453 |

Fig. 8. Voltage Threshold Adaptive Memristor (VTEAM) framework for fitting the I-V hysteresis of the n-GaAs/$Al_2O_3$/p-Si memristor heterojunction.

As illustrated in Fig. 8, The fitted VTEAM parameters reveal physically meaningful distinctions between the four measurement sweeps and reflect the underlying conduction mechanisms of the n-GaAs/$Al_2O_3$/p-Si heterojunction. For the high-resistance negative-bias sweep (blue, $0 \rightarrow -8$ V), the device begins in the fully depleted OFF state, reflected in the initial state variable fraction $w_{init}/w_{off} \approx 0.99$. The fitted negative threshold voltage $v_{on} \approx -5.4$ V indicates that significant state evolution - corresponding to partial filament formation or interface trap filling - commences only at elevated reverse bias, consistent with the observed monotonic resistance reduction from 35.5 kΩ at $-0.1$ V to 21.5 kΩ at $-6.0$ V. The relatively small magnitude of $k_{on}$ ($\approx 10^{-5}$ m/s) reflects the sluggish, thermally-activated nature of this ion migration or defect redistribution process under moderate bias. The abrupt discontinuity at $-6.1$ V, where resistance drops sharply from 21.5 to 17.8 kΩ - a 1.21× step - marks the SET event, likely corresponding to a percolation threshold in the conductive filament. This transition is reproduced by the VTEAM state variable reaching $w_{on}$, i.e., full switching. The higher residual log-RMS error for this curve (0.48) reflects the fundamental limitation of a single-threshold, two-state VTEAM formulation in simultaneously capturing the gradual pre-SET resistance drift and the abrupt switching discontinuity, which in reality arise from distinct physical processes (bulk trap filling versus localized filament snap-through). For the low-resistance return sweep (orange, $-8 \rightarrow 0$ V), the device is already in the ON state at the start of the sweep ($w_{init}/w_{off} \approx 0.99$ toward $w_{on}$), reflecting that the SET event occurred during the prior forward sweep. The orange sweep exhibits a nearly constant resistance of approximately 6.4 – 7.1 kΩ across the measured window (− 5.8 to − 0.5 V), representing a 3.4 × lower resistance than the blue curve at the same voltage, consistent with a stable, low-resistance conductive filament. The excellent fit quality (log-RMS = 0.039) confirms that the VTEAM ohmic model with a fixed effective resistance accurately represents the purely resistive, filament-dominated transport in this state. The extrapolated model line - extended to 0 V - shows the expected smooth current decay to zero as the voltage is removed, with no evidence of spontaneous RESET, indicating a non-volatile ON state under zero bias. For the forward-bias sweeps (red and green), a diode-shaped current–voltage relationship was required, with the fitted saturation current $I_0 \sim 10^{-4} - 10^{-3}$ A and ideality factors $n \sim 140 - 150$. The large ideality factors, substantially exceeding the ideal diode value of unity, are characteristic of generation–recombination current or tunnelling-dominated transport through the $Al_2O_3$ interlayer. The red sweep (0 → + 4 V) and green sweep (+ 4 → 0 V) show a modest but reproducible current asymmetry at + 4 V (148.6 μA versus 174.1 μA), with the green curve carrying ~ 17% more current — consistent with a slightly enhanced conduction path established during the negative-bias SET cycle that persists into the forward direction. The near-zero $w_{init}$ for the green curve ($w_{init}/w_{off} \approx 0.19$) confirms that the device enters the positive return sweep in a partially switched state. The positive threshold $v_{off}$ fitted to approximately 3.5 – 12.5 V lies outside the measurement window for all forward sweeps, correctly predicting that no positive-bias switching event is triggered within the 0 – 4 V measurement range, consistent with the smooth, hysteresis-free character of the red curve.